\documentstyle[prl,aps,epsf]{revtex}
\begin{document}
\draft

\twocolumn[\hsize\textwidth\columnwidth\hsize\csname@twocolumnfalse\endcsname

\title{Mode-locking in ac-driven vortex lattices with random pinning}

\author{Alejandro B. Kolton and Daniel Dom\'{\i}nguez}
\address{Centro At\'{o}mico Bariloche, 8400 S. C. de Bariloche,
R\'{\i}o Negro, Argentina}
\author{Niels Gr{\o}nbech-Jensen}
\address{Department of Applied Science, University of California, Davis, CA
95616, USA\\
NERSC, Lawrence Berkeley National Laboratory, Berkeley, CA 94720, USA}

\date{\today}
\maketitle
\begin{abstract}
We find
mode-locking steps in simulated current-voltage characteristics of
ac-driven vortex lattices with {\it random} pinning. For low frequencies there
is mode-locking above a finite ac force amplitude, while for large
frequencies there is mode-locking for any small ac force.
This is correlated with the nature of temporal order in the different
regimes in the absence of ac drive.
The mode-locked state is a frozen solid pinned in the moving reference
of frame, and the depinning from the step shows plastic flow and
hysteresis.

\end{abstract}


]                

\narrowtext

In 1971 A. T. Fiory \cite{fiory} observed
steps in the current-voltage (IV)
characteristics of ac-driven superconducting thin films,
analogous to the Shapiro steps found in Josephson junctions
\cite{shapiro}.
The steps were observed for voltages such that 
$2\pi \langle v\rangle /a_0 = (p/q) \Omega$, with 
$\langle v\rangle$ the average vortex velocity, 
$a_0$ the triangular vortex lattice period,
$\Omega$ the frequency of the external ac-drive, and $p,q$
integers. This is a particular case of mode-locking,
where an internal frequency of the system (in this case 
$\omega_0=\langle v\rangle 2\pi/a_0$)
locks to a rational multiple of the external frequency.
Several other systems with many degrees of freedom, such as 
charge density waves \cite{cdw,cdw2}, spin density waves \cite{sdw}, 
Josephson junction arrays\cite{jja} 
and superconductors with periodic pinning arrays \cite{periexp,charles}, 
also exhibit mode-locking behavior experimentally.
Recently, Harris {\it et al.} \cite{harris}
observed the Fiory steps in YBCO and found that they
vanish when the vortex melting line is crossed.
Historically, the experiment of Fiory motivated the landmark 
works of Schmid and Hauger \cite{SH} and Larkin and Ovchinikov \cite{LO} 
on the dynamics of moving vortex lattices. In \cite{SH,LO} it was assumed
that at large velocities vortices form a perfect triangular lattice
in which the effect of random pinning could be  treated perturbatively.
However, it is now clear that, instead of a perfect triangular
lattice, there are several other dynamical phases of driven vortices: 
plastic flow \cite{plastic}, 
moving smectic, transverse moving glass and moving Bragg glass
\cite{KV,GLD,BMR}. 
These regimes have been observed experimentally \cite{EXP} and 
in numerical studies \cite{SIM,kolton}. Thus, it may be of interest to study 
how the Fiory steps can arise in different moving vortex phases.
Moreover, a peak 
in the voltage noise power spectrum at the ``washboard''
frequency $\omega_0$ has recently been observed experimentally in driven
superconductors \cite{troya,togawa}. This is a signature of 
temporal order in the high-current  steady states of moving
vortices \cite{BMR,BF}. We here present a
numerical study of
mode-locking for the different regimes of vortex velocities and its
relationship with temporal order.

The equation of
motion of a vortex in position ${\bf r}_i$ is:
\begin{equation}
\eta \frac{d{\bf r}_i}{dt} = -\sum_{j\not= i}{\bf\nabla}_i U_v(r_{ij})
-\sum_p{\bf \nabla}_i U_p(r_{ip}) + {\bf F}(t),
\end{equation} 
where $r_{ij}=|{\bf r}_i-{\bf r}_j|$ is the distance between vortices $i,j$,
$r_{ip}=|{\bf r}_i-{\bf r}_p|$ is the distance between the vortex $i$ and
a pinning site at ${\bf r}_p$, $\eta=\frac{\Phi_0H_{c2}d}{c^2\rho_n}$ is the
Bardeen-Stephen friction and ${\bf F}(t)=
\frac{d\Phi_0}{c}[{\bf J_{dc}}+{\bf J}_{ac}\cos(\Omega t)]\times{\bf z}$
is the driving force due to an alternating current ${\bf J}_{ac}\cos(\Omega t)$
superimposed to a constant  current ${\bf J}_{dc}$.
A 2D thin film superconductor of thickness
$d$ with $d\ll\lambda$, has an effective
penetration depth $\Lambda=2\lambda^2/d$. Since $\Lambda$ is of the order
of the sample size, the vortex-vortex interaction is considered logarithmic:
$U_v(r)=-A_v\ln(r/\Lambda)$, with $A_v=\Phi_0^2/8\pi\Lambda$
\cite{kolton}.
The vortices interact with a random distribution of
attractive pinning centers with 
$U_p(r)=-A_p e^{-(r/\xi)^2}$, $\xi$ being the coherence length. 
Length is normalized by $\xi$, energy by $A_v$, 
and time by 
$\tau=\eta\xi^2/A_v$.  We consider $N_v$ vortices and $N_p$ pinning
centers in a rectangular box of size $L_x\times L_y$, 
and the normalized  vortex density is $n_v=N_v\xi^2/L_xL_y=B\xi^2/\Phi_0$.
Moving vortices induce a total electric field  ${\bf
E}=\frac{B}{c}{\bf v}\times{\bf z}$, with ${\bf v}=\frac{1}{N_v}\sum_i 
{\bf v}_i$. 

We study the response of the vortex lattice to an external ac+dc 
force of the form
${\bf F}=[F_{dc}+ F_{ac}\cos(\Omega t)]{\bf y}$ \cite{trans_ac} at 
$T=0$, solving Eq.~(1) for different values of $F_{ac}$ and $\Omega$.
The simulations are for constant vortex density $n_v=0.04$ in
a box with $L_x/L_y=\sqrt{3}/2$, and 
$N_v=64,100,144,196,256,400$ (we show results for $N_v=256$), and
we consider weak pinning strength of $A_p/A_v=0.05$
with a density of pinning centers being $n_p=0.08$. Periodic boundary
conditions are imposed on the simulation box and the resulting long-range
interaction is determined by Ref.~\cite{log}.
The equations are integrated
using a time step of $\Delta t=0.001\tau$, averages are
evaluated during $131072$ steps after $3000$ steps for 
equilibration (when the total energy reaches a stationary mean value). 

Let us first review the behavior for $F_{ac}=0$. There are three
dynamical regimes when increasing $F_{dc}$ above  the critical
depinning force, $F_c$, in the case $F_{ac}=0$: plastic flow  for $F_c <
F_{dc}< F_p$, smectic flow for 

\begin{figure}
\centerline{\epsfxsize=8.5cm \epsfbox{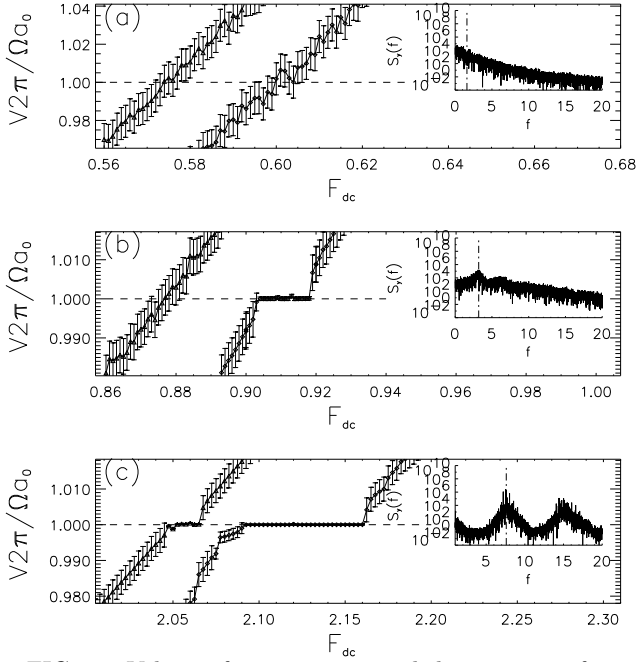}}
\caption{
Velocity-force curve  around the main
interference condition $V = \Omega a_0/2 \pi$ for three typical 
drive frequencies $\Omega$. Each case show results 
for two values of amplitude $F_{ac}$ 
(the curves are shifted in $F_{dc}$ for clarity). 
Insets show corresponding voltage power spectrum for 
$F_{ac}=0$ and $V \approx V_{step}$. Vertical dashed line in the 
spectral density indicates the washboard frequency.
(a) $\Omega=0.5$, $F_{ac}=0.4$ (left), $F_{ac}=1.8$ 
(right). 
(b) $\Omega= 1$, $F_{ac}=0.4$ (left), 
$F_{ac}=3$ (right). 
(c) $\Omega = 2.5$, $F_{ac}=0.75$ (left), 
$F_{ac}=4$ (right).}
\end{figure}\noindent

$F_p < F_{dc}< F_t$, and a transverse
solid  for $F_t < F_{dc}$ (see \cite{kolton}). 
 The characteristic
forces in our case are  $F_c \approx 0.15$,  $F_p \approx 0.6$ and
$F_t\approx1.2$. Since the nature of mode-locking is related to the
existence of temporal order at the  washboard frequency $\omega_0$ we
analyze the voltage noise $S_y(f)=|\frac{1}{T}\int_0^Tdt 
(V_y(t)-V)\exp(i2\pi ft)|^2$ in each dynamical regime. In  the insets
of Fig.~1(a-c) we show the spectral densities  corresponding to each
regime and we  indicate the corresponding $\omega_0$.  In the inset of
Fig.~1(a) we see that there is no temporal order at the washboard
frequency, and only the typical broad band noise of plastic flow is
observed\cite{plastic,togawa}.  For the smectic flow regime, we see in
the inset of Fig.~1(b) that there is  a small and broad peak at
$\omega_0$.  Only for large forces, $F>F_t$, in the transverse solid
regime well developed peaks appear at the washboard frequency
\cite{togawa}  and harmonics (see the inset of Fig.~1(c)). 

We now 
study the response of each of these dynamical regimes 
with velocity $V=V(F_{dc})$ 
to the superimposed ac-force $F_{ac}\cos(\Omega t)$, for 
varying values of $F_{ac}$ for a given $\Omega$.
We expect the main interference step ($p=q=1$) to occur when
$V=V_{step}=\Omega a/2 \pi$  ({\it i.e.}, $\Omega=\omega_0$) if there
is mode-locking. 
We therefore choose the values of $\Omega$ 
such that the expected step, $V_{step} = \Omega a/2\pi$,
would correspond to velocities $V$ belonging 
to a given dynamical regime 

\begin{figure}
\centerline{\epsfxsize=8.7cm \epsfbox{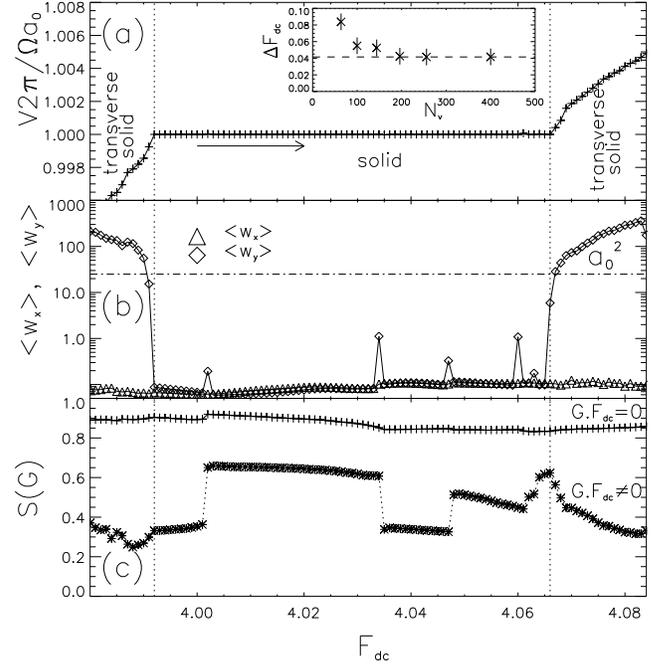}}
\caption{(a) Velocity-force curve  around the main
interference step for $\Omega=5$ and $F_{ac}=6$. Inset shows the
finite size dependence of the step width. 
(b) Time averaged quadratic mean displacements in the longitudinal direction
$\langle w_y(t) \rangle$, ($\Diamond$) points, and in the 
transverse direction $\langle w_x(t) \rangle$, ($\triangle$) points. 
The dash-dotted line indicates $a_0^2$.  
(c) Intensity of the Bragg peaks. For smectic ordering $S(G_1)$, $K_y=0$: ($+$)
points. For longitudinal ordering $S(G_{2,3})$, $K_y=0$: ($*$) points.}
\end{figure}\noindent 

of the limit $F_{ac}=0$. Each simulation is started at
$\langle v_y \rangle \approx 0.975 \Omega a/2 \pi $ 
with an ordered triangular lattice 
up to values such that 
$\langle v_y \rangle \approx 1.025 \Omega a/2 \pi$ 
by slowly increasing the dc force $F_{dc}$ with $\Delta F_{dc}= 0.001$. 
For low $\Omega$, for which
we have plastic flow when  $F_{ac}\rightarrow0$, we
find that there are no interference steps in a wide range of  $F_{ac}$
(shown in Fig.~1(a) for $F_{ac}/V_{step}<1$ (left curve) and 
$F_{ac}/V_{step}>1$ (right curve)).
This is consistent with the 
observed lack of temporal order  in the inset 
of Fig.~1(a) and indicates that very large $F_{ac}$ 
are possibly needed in order to induce mode-locking.
For intermediate $\Omega$, for which  we
have smectic  flow when $F_{ac}\rightarrow0$, 
we find that there are no steps
for small amplitudes, $F_{ac}/V_{step}<1$, while there are steps for
$F_{ac}/V_{step}>1$, as shown in Fig.~1(b) in  the left and right curves,
respectively. 
This means that the small washboard peak observed in the inset of
Fig.~1(b) is not large enough to induce mode-locking steps
for small $F_{ac}$. However, this short-range temporal order 
can be amplified  for intermediate values of $F_{ac}$ 
giving place to steps in this case.
For high $\Omega$, corresponding  to a  transverse solid regime 
when $F_{ac}\rightarrow0$  we find that  there are steps both for small
$F_{ac}/V_{step}<1$ and large $F_{ac}/V_{step}>1$ values of the  ac
amplitude, as we can observe in Fig.~1(c).
This is in agreement with the $F_{ac}=0$ spectral response 

\begin{figure}
\centerline{\epsfxsize=8.5cm \epsfbox{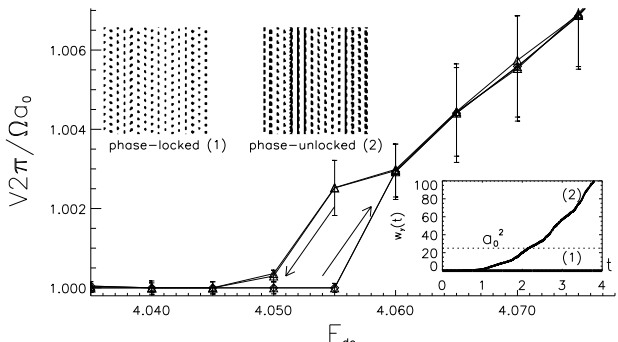}}
\caption{ Velocity-force curve around
the depinning from the main step for $\Omega=5$ and $F_{ac}=6$.
The upper left insets show typical time averaged coarse-grained 
density of vortices 
seen from a system of reference moving with velocity 
$v_{step}=\Omega a_0/2\pi$, 
for a mode-locked state $F_{dc}=4.045$ (left) and for an unlocked state
$F_{dc}=4.085$ (right). The lower right inset shows the corresponding
typical quadratic mean longitudinal displacements for both cases. Dashed line 
indicates $a_0^2$.}
\end{figure}\noindent

observed in  
inset of Fig.~1(c). In this case, the temporal
order is robust enough for mode-locking to be produced by small values of
$F_{ac}$.

Let us now examine in detail the dynamics and the structural order
within and in the vicinity of a mode-locked step, in the transverse
solid case.
In Fig.~2(a) we show a typical $V-F_{dc}$ curve around the step.  
In the inset of Fig.~2(a)
we show a finite size analysis of the step width for  $N_v=64,
100, 144, 196, 256, 400$, where the error bars are due to the observed 
dependence of the width in three different realizations of disorder. We
observe that for $N_v > 256$ the step width tends 
to saturate in a size-independent value. To
analyze  the dynamical behavior we define  the
quadratic mean displacements of vortices in directions parallel
$w_y(t)$ and perpendicular  $w_x(t)$  to the external force,
calculated from the center of mass position  $(X_{cm}(t),Y_{cm}(t))$ as: 
$w_x(t)=\frac{1}{Nv}\sum_i[\tilde{x}_i(t)-\tilde{x}_i(0)]^2$ 
and $w_y(t)=\frac{1}{Nv}\sum_i[\tilde{y}_i(t)-\tilde{y}_i(0)]^2$,
where  $\tilde{x}_i(t)=x_i(t)-X_{cm}(t)$ y
$\tilde{y}_i(t)=y_i(t)-Y_{cm}(t)$. In   Fig.~2(b) we show the time
average of these quantities, $\langle w_x(t) \rangle$ and  $\langle
w_y(t) \rangle$, as a function of $F_{dc}$. Outside the step we observe
that the transverse mean displacement (TMD) is limited 
$\langle w_x(t) \rangle \ll a_0^2$, while the  
longitudinal mean displacement (LMD) is unbounded 
$\langle w_y(t) \rangle \gg a_0^2$. 
This corresponds to a state with only longitudinal 
diffusion (i.e., a transverse solid \cite{kolton}). 
Noticeably, in the
transition to the synchronization, the LMD freezes  in a value,
$\langle w_y(t) \rangle \ll a_0^2$, while the TMD
remains  practically constant. This {\it mode-locking
longitudinal freezing} 
can also be  observed as a dramatic decrease of
the low frequency voltage noise in both directions.   
The mode-locked state is therefore a frozen solid.  
To study the translational order we calculate
the structure  factor  as $S({\bf k})=
\langle|\frac{1}{N_v}\sum_i \exp[i{\bf k}\cdot{\bf r}_i(t)]|^2\rangle$.
In this regime there are smectic order peaks of 
magnitude $S_s = S({\bf G}_1)$ with 

\begin{figure}
\centerline{\epsfxsize=8.5cm \epsfbox{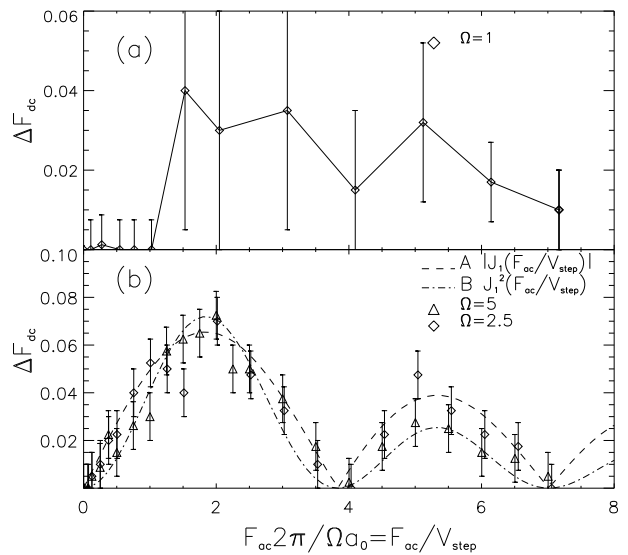}}
\caption{ Step width $\Delta F_{dc}$ vs $F_{ac} 2\pi/\Omega a_0=F_{ac}/
V_{step}$. (a) $\Omega = 1$. (b) $\Omega= 5$ 
($\triangle$) points and $\Omega= 2.5$ ($\Diamond$) points. Dashed line
shows a fit to  $A|J_1(F_{ac}/V_{step})|$ and
the dot-dashed line a fit to $B|J_1(F_{ac}/V_{step})|^2$.}  
\end{figure}\noindent

${\bf G}_1=(\pm2\pi/a_0,0)$ and longitudinal peaks of magnitude 
$S_l=(S({\bf G}_2)+S({\bf G}_3))/2$ with ${\bf G}_2=\pm 2\pi/a_0(1/2,
\sqrt{3}/2)$ and ${\bf G}_3=\pm2\pi/a_0(-1/2, \sqrt{3}/2)$. In Fig.~2(c)
we plot  $S_s$ and $S_l$. We do not observe any important change
in the  transition to the mode-locked state. Inside the steps we
observe jumps  in $S_l$ indicating that there are different metastable 
mode-locked structures. In Fig.~3 we
analyze in detail the process of depinning from  the step. 
We show a detailed view of the $V-F_{dc}$ curve in the transition from the 
synchronized regime. Varying $F_{dc}$ back and forth, we observe a
clear hysteresis cycle. In the lower right inset of Fig.~3 we show the
evolution of $w_y(t)$ with time inside the step and
outside the step. While $w_y(t) \ll a_0$ for all $t$
inside the step, outside there is ballistic diffusion $w_y(t) \sim
t^2$. To visualize the spatial structure in the transition we
define a coarse-grained vortex density  $\rho'_v({\bf r},t)$ seen from
a system of reference moving with velocity $V_{step}$  as follows:
$\rho'_v({\bf r},t)=\frac{1}{N_v} \sum_i \delta({\bf r}-{\bf
r'}_i(t))$, where ${\bf r'}_i(t)={\bf r}_i(t)-{\bf y} V_{step}t$. We
take a coarse-graining  scale $\Delta r < a_0$. In the upper left insets 
of Fig.~3 we show
the temporal average  $\langle \rho'_v({\bf r},t) \rangle$ of the
density, inside (mode-locked) and outside (mode-unlocked) the step. This quantity shows
the stationary trajectories of vortices seen from a  moving frame of
reference with velocity $V_{step}$. Inside the step we see 
that the vortices are localized, oscillating around their equilibrium
positions in a moving lattice with  velocity $V_{step}$. This is
in agreement with the mode-locked frozen solid inferred 
from the result  shown in Fig.~2(b), which is ``pinned'' in 
the moving reference frame with velocity $V_{step}$. 
Just above the depinning from the step we see that some vortices delocalize  
following straight trajectories parallel to the force
around ``pinned'' vortices,  producing coexistence of mode-locked and
mode-unlocked channels of flow.  This could be interpreted
as a one-dimensional ``plastic'' depinning from the step.

Mode-locking of the 
steps in the $V-F_{dc}$ curve can be characterized qualitatively by
how the dc current range in mode-locking depends on
$F_{ac}$ and $\Omega$, (as
it  was done for example in \cite{jja,periexp,charles}). 
In Fig.~4(a-b) we show the range (width)
$\Delta F_{dc}$ for the case  $F_p < F_{dc} < F_t$ and $F_t < F_{dc}$
respectively. The error bars and the mean values  were estimated by
repeating the simulation for three different disorder realizations. In
Fig.~4(a) we show $\Delta F_{dc}$ for $\Omega=1$ vs $F_{ac}$, which
corresponds to the  smectic flow regime for $F_{ac}\rightarrow0$. 
We see that there
is mode-locking  above  a finite critical value
$F_{ac}/V_{step} \approx 1$ (see Fig.~1(b)). For larger  amplitudes we
were not able to obtain a systematic dependence on amplitude because 
the step widths depend strongly
on the disorder realization in this case. 
In Fig.~4(b) we show  $\Delta F_{dc}$ for two frequencies 
$\Omega=2.5, 5$ vs $F_{ac}$,  
which correspond  to the transverse solid in the
$F_{ac}=0$ limit. We can collapse (approximately) both
curves into a single curve if we plot $\Delta F_{dc}$  vs
$F_{ac}/V_{step}$. A dependence, $\Delta F_{dc} \sim (1/C_{66})
|J_1(F_{ac}/V_{step})|^2$, was found by Schmid and Hauger \cite{SH} 
(and also found in other elastic models) where $V_{step}=a_0 \Omega/2 \pi$
and $C_{66}$ is the shear modulus. This result was obtained
using a perturbative approach in an  elastic
model for the vortex lattice.
Strikingly, our results seem to follow more closely a dependence
of the form $\Delta F_{dc} \approx A
|J_1(F_{ac}/V_{step})|$   with $A$ being a
constant.  This dependence is the same
found for the one-dimensional problem of an overdamped
single Josephson  junction
or a particle moving in a periodic potential. 
In our case, 
a linear dependence  of the mode-locking intensity with $F_{ac}$ would 
be a consequence of the existence of temporal order in the
$F_{ac}=0$ limit. This was not taken into account in the 
perturbative calculation of
Schmid and Hauger where   
mode-locking arises as a second order effect.

In conclusion, it is possible to have
mode-locking in driven vortices with {\it random} pinning for high
enough frequencies,  in agreement with 
the experiments of Fiory \cite{fiory} and Harris {\it et al}
\cite{harris}. The mode-locked state can be viewed as a frozen solid
pinned in the moving frame of reference, and the depinning from mode-locking
is plastic and hysteretic. Also, the response to an ac drive for 
different frequencies can be an interesting experimental probe of the
dynamical regimes of driven vortices.

We acknowledge discussions with L.~Balents, P.~S.~Cornaglia, M.~F.~Laguna, 
V.~I.~Marconi. This work has been supported by 
CONICET, Fundaci\'{o}n Antorchas and ANPCYT (Argentina) and by
the Director, Office of
Advanced Scientific Computing Research, Division of Mathematical, Information
and Computational Sciences of the U.S.~Department of Energy under
contract number DE-AC03-76SF00098.

\end{document}